\documentclass[a4paper,10pt]{article}

\usepackage{amssymb,amsmath} 
\usepackage{bm} 

\def\eq#1{{Eq.~(\ref{#1})}}

\def\eq#1{{Eq.~(\ref{#1})}}

\def\pdov#1#2{\frac{\partial#1}{ \partial#2}}

\def\gu#1#2{{g^{#1#2}}}
\def\gl#1#2{{g_{#1#2}}}
\def\g{\sqrt{-g}\,}

\newcommand{\w}[1]{\bm{#1}}
\newcommand{\el}{\w{\ell}}
\newcommand{\we}{\w{e}}
\newcommand{\LL}{Lanczos-Lovelock}

\newcommand{\Cal}[1]{\ensuremath{\mathcal{#1}}}
\newcommand{\ph}[1]{\phantom{#1}}
\newcommand{\D}{\ensuremath{\nabla}}

\newcommand{\mdof}{microscopic degrees of freedom}

\def\g{\sqrt{-g}\,}
\def\gu#1#2{{g^{#1#2}}}
\def\gl#1#2{{g_{#1#2}}}
        
\def\pdov#1#2{\frac{\partial#1}{ \partial#2}}

\def\ch#1#2{{\chi^{#1}_{\phantom{#1}#2}}}    
\def\pip#1#2{{\Pi^{#1}_{\phantom{#1}#2}}}    

\title{Entropy density of spacetime and the Navier-Stokes fluid dynamics of null surfaces}
\author{T Padmanabhan\\
IUCAA, Pune, INDIA 411 007\\
email: paddy@iucaa.ernet.in}

\date{  }
\begin{document}

\maketitle

\begin{abstract}
It has been known for several decades that Einstein's field equations, when projected onto a null surface, exhibits a structure very similar to non-relativistic Navier-Stokes equation. I show that this result arises quite naturally when gravitational dynamics is viewed as an emergent phenomenon. Extremising the spacetime entropy density associated with the null surfaces leads to a set of equations which, when viewed in the local inertial frame, becomes  identical to the Navier-Stokes (NS) equation.  This is in contrast to  the usual description of Damour-Navier-Stokes (DNS) equation in a general coordinate system, in which  there appears a Lie derivative rather than convective derivative. 
I discuss this difference, its importance and why it is more appropriate to view the equation in a local inertial frame. The viscous force on fluid, arising from the gradient of the viscous stress-tensor, involves the second derivatives of the metric and does not vanish in the local inertial frame while the viscous stress-tensor itself vanishes  so that inertial observers detect no dissipation.
 We thus provide an entropy extremisation principle that leads  to the DNS equation, which makes the hydrodynamical analogy with gravity completely natural and obvious. Several implications of these results are discussed.
\end{abstract}

\section{Gravity as an emergent phenomenon}

Large amount of theoretical evidence suggests that gravity could be an emergent phenomenon like gas dynamics or elasticity
with the gravitational field equations having the same status as, say, the equations of gas/fluid dynamics \cite{rop}. They seem to describe the thermodynamic limit of the statistical mechanics of (as yet unknown) atoms of spacetime or, equivalently, hydrodynamic (long wavelength) limit of a suitably defined `fluid' system. 

A very direct argument in favour of this paradigm is the validity of principle of equipartition which allows one to use the classical field equations, along with the expression for Davies-Unruh temperature \cite{daviesunruh}, to determine the number density of microscopic degrees of freedom \cite{surfaceprd}. One finds that any null surface in the spacetime is endowed with a number density of \mdof\ (`atoms of spacetime') which depends on the structure of the gravitational theory. In the simplest context of Einstein's theory, this density is a constant (equal to one degree of freedom per Planck area) while in a generic \LL\ model, it is proportional to $P^{abcd}=\partial L/\partial R_{abcd}$ where $L$ is the \LL\ Lagrangian and $R_{abcd}$ is the curvature tensor \cite{surfaceprd}. We shall hereafter call $P^{abcd}$ as the \textit{spacetime entropy tensor} or simply the \textit{entropy tensor} of the theory. (In the case of Einstein's theory with $L\propto R$, we have $P^{ab}_{cd}\propto (\delta^a_c\delta^b_d-\delta^a_d\delta^b_c)$).
Given the number density of \mdof\ one can construct an entropy functional for the system which will depend only on $P^{abcd}$ and  lead correctly to the  Wald entropy \cite{wald} of the \LL\ theory. The crucial fact that entropy of a horizon in a gravitational theory depends on the theory (and is proportional to the area \textit{only} in the simple case of Einstein gravity) finds expression in the related fact that the density of \mdof\, determined through the law of equipartition, depends on the theory. 

This key  `internal evidence' from the structure of gravitational theories --- especially in \LL\ models --- suggests that they may be viewed as the thermodynamic limit of the statistical mechanics of the \mdof. This is reminiscent of the work by Boltzmann and others, who used the `internal evidence' of thermal phenomena to \textit{deduce the existence} of atoms/molecules and  determine the Avogadro number which quantifies their  number density even without knowing what they are and having no direct observational support for their existence. 

The above results were obtained by starting from the field equations of the theory, rewriting them in the form of law of equipartition and thus determining the density of \mdof. A more attractive procedure will be to invert this argument and obtain the field equations of the theory from the density of \mdof. We do know that the thermodynamical behaviour of any system can be described by an extremum principle for a suitable  potential (entropy, free energy ...) treated as a functional of appropriate variables (volume, temperature ,....). If our ideas related to gravitational theories are correct, it must be possible to obtain the field equations by extremising a suitably defined entropy functional. The fact that null surfaces block information suggests that the entropy functional should be closely related to null surfaces in the spacetime. This turns out to be true. The \LL\ field equations can indeed be obtained \cite{aseemtp} by extremising an entropy functional associated with the null vectors in the spacetime. This functional depends only on a tensor $P^{abcd}$ which has the symmetries of the curvature tensor and is divergence-free in all indices. One can show that the thermodynamic variational principle leads to the  field equations of a \LL\ model with the same entropy tensor. 

In this paper, I will reinterpret these results in a different manner. I will argue that it is more natural to project the  equations which result from the entropy extremisation onto the null surfaces of the spacetime, thereby obtaining what is usually called in the literature as the Damour-Navier-Stokes (DNS hereafter) equation \cite{damourthesis, damourGR}. In general this equation is \textit{similar to} Navier-Stokes (NS) equation but is \textit{not} identical to it because the DNS equation contains a Lie derivative while the standard NS equation has a convective derivative. (As I will describe in Section \ref{sec:DNS}, this structural difference exists in all the previous derivations of DNS equation in the literature but has not been emphasized). I will argue that it is better to study the structure of gravitational field equations in freely falling frames rather than in arbitrary coordinates and will show that --- in locally inertial frames --- the DNS equation becomes \textit{identical} to NS equation. Moreover, the physical interpretation, especially aspects related to viscous dissipation, becomes transparent in this frame.

\textit{We thus obtain an entropy extremisation principle to derive DNS equation directly, which makes the hydrodynamical analogy with gravity self contained.} This is in contrast with the conventional approach in which one first obtains the field equations from some (action) principle and then interprets it as a NS like equation. In  such an approach it seems somewhat mysterious that the equation looks similar to NS equation with a thermodynamic structure. 

This paper is organized as follows: I begin in the next section with a short review of the derivation of gravitational field equations  by extremising the entropy density of spacetime. In Section
\ref{sec:entmaxproj}, I argue in favour of projecting these equations onto a null surface.
The technology to describe the extrinsic geometry of the null surface is reviewed briefly in in Section \ref{sec:nulltech}. (A more pedagogical discussion is included in the Appendix to make the paper self-contained.) This is used to recast the equations in the form of NS equation, with convective derivative, in a boosted inertial frame in Section \ref{sec:NSinboost}. The DNS equation in a general frame (which has the Lie derivative rather than convective derivative) is obtained in Section \ref{sec:DNS} and a comparison of these equations and their physical implications are presented in Section \ref{sec:DNSdisc}. Some further technical issues related to rescaling of null normal and introducing spacetime dependent boosts are discussed in Section \ref{sec:rescale} and the conclusions are presented in the last section.
This analysis also opens up several further avenues of research which I will briefly mention whenever appropriate.

 We will use the mostly positive signature; the 
Latin indices, $a,b...$ go over $0-3$ in the spacetime manifold $\mathcal{M}$ while the Greek indices $\alpha,\beta, ...$ go over the  coordinates in a 3-dimensional null surface $\mathcal{S}$ with signature $(-, +, +)$. When we restrict ourselves to the two-dimensional, spatial sub-manifold of this null surface, we will use uppercase Latin indices $A,B,...$.

\section{Entropy density of spacetime and its extremisation}

The field equations of \LL\ models were obtained previously \cite{rop, aseemtp} from an entropy extremisation principle along the following lines, which we shall rapidly review. (The  physical motivation and algebraic details are described in detail in previous papers, e.g., ref.\cite{rop}, and will not be repeated here.) The procedure involves associating with every null vector in the spacetime an entropy functional  $S_{grav}$ and demanding
$\delta[S_{grav}+S_{matter}]=0$ for \textit{all null vectors in the spacetime} where $S_{matter}$ is the relevant  matter entropy. It can be argued, based on the energy flux across local Rindler horizons and the associated entropy, that the 
 form of $S_{matter}$  \textit{relevant for this purpose} can be taken to be 
\begin{equation}
S_{\rm matt}=\int_\mathcal{V}{d^kx\sqrt{-g}} \, T_{ab}n^an^b
      \label{Smatt}
\end{equation} 
where $n^a$ is a null vector field and $T_{ab}$ is the matter energy-momentum tensor living in a general $D(\ge 4)$ dimensional spacetime.  (The integration could be over this spacetime volume ($k=D$) or even over a nontrivial sub-manifold $(k<D$), say, a set of null surfaces. This does not affect the variational principle or the resulting equations.) The simplest choice for $S_{grav}$ is a quadratic expression  in the derivatives of the  null vector:  
 \begin{equation}
S_{grav}= - 4\int_\mathcal{V}{d^kx\sqrt{-g}}\, P_{ab}^{\ph{a}\ph{b}cd} \D_cn^a\D_dn^b 
    \label{Sgrav}
\end{equation}  
where  $P_{ab}^{\ph{a}\ph{b}cd}$ is a tensor having the symmetries of curvature tensor and is divergence-free in all its indices.
(This makes the notation $P_{ab}^{cd}\equiv P_{ab}^{\ph{a}\ph{b}cd}$ unambiguous and we will often use this placement of indices.) It can be shown that the most general tensor which satisfies these criteria is the entropy tensor of a \LL\ theory; that is, the $P^{abcd}$ in \eq{Sgrav}   can be expressed as  $P^{abcd}=\partial L/\partial R_{abcd}$ where $L$ is the \LL\ Lagrangian and $R_{abcd}$ is the curvature tensor \cite{rop}. 
This choice  will also ensure that the  equations resulting from the entropy extremisation do not contain any derivative of the metric which is of higher order than second. (More general possibilities exist which we will not discuss in this paper.)  The expression for the total entropy  now becomes:
\begin{equation}
S[n^a]=-\int_\Cal{V}{d^kx\sqrt{-g}} \left(4P_{ab}^{cd} \D_cn^a\D_dn^b -  T_{ab}n^an^b\right) \,,
\label{ent-func-2}
\end{equation}
 We now vary the  vector field $ n^a$  in \eq{ent-func-2} after adding a
Lagrange multiplier function $\lambda(x)$ for imposing the   condition
$ n_a\delta  n^a=0$. Straight forward algebra  (see e.g., Section 7.1 of ref. \cite{rop}) now shows that the condition $\delta[S_{grav}+S_{matter}]=0$ leads to the equations
\begin{equation}
\left(2\mathcal{R}^a_b -  T{}^a_b+\lambda \delta^a_b\right) n_a=0\,, 
\label{ent-func-6}
\end{equation}
where $\mathcal{R}^a_b\equiv P_b^{\ph{b}ijk}R^a_{\ph{a}ijk}$ is the generalization of Ricci tensor to \LL\ models.
We  demand that this  extremum condition should hold for all 
null vector fields
$ n^a$. 

At this stage the usual procedure \cite{rop,aseemtp} is to use the
generalized Bianchi identity and the condition $\nabla_aT^a_b=0$ to obtain from \eq{ent-func-6} the  equations
\begin{equation}
\mathcal{G}^a_b =  \mathcal{R}^a_b-\frac{1}{2}\delta^a_b L = \frac{1}{2}T{}_b^a +\Lambda\delta^a_b   
\label{ent-func-71}
\end{equation}
where $\Lambda$ is a constant and  $P^{abcd}=\partial L/\partial R_{abcd}$. These are  precisely the field equations for  gravity  in a theory with \LL\ Lagrangian $L$. When $L\propto R$, leading to $P^{ab}_{cd}\propto (\delta^a_c\delta^b_d-\delta^a_d\delta^b_c)$,
we have $\mathcal{R}^a_b \propto R^a_b, \mathcal{G}^a_b \propto G^a_b$ and one recovers Einstein's equations.

There exists, however an alternative route which we will explore in this paper, that turns out to be more in tune with the entropy extremisation principle and emergent perspective of gravity. To do this we note that, while \eq{ent-func-6} holds for any vector field, once the normalization condition is imposed through the Lagrange multiplier, the entropy was originally attributed to \textit{null} vectors and hence it is natural to study  \eq{ent-func-6} when $n^a$ is   the null normal $\ell^a$ of a null surface $\mathcal{S}$ in the spacetime and project 
\eq{ent-func-6} onto the null surface. This is what we will do.

In the case of spacelike or timelike surfaces with a normal, such a projection is straightforward but the null surfaces require a somewhat more careful treatment. This is described  in the Appendix  for those readers who may not be completely familiar with the issues involved. In the main text of the paper, we shall use the results from this Appendix  to proceed further. We will also work with Einstein's theory in $D=4$ for clarity and definiteness and will mention more general possibilities in the end.

\section{Navier-Stokes dynamics of null surfaces}

\subsection{Entropy extremum condition projected onto a null surface}
\label{sec:entmaxproj}

In a 4-dimensional spacetime manifold $\mathcal{M}$ with a metric $g_{ab}$, a  
null surface  with normal $\el$ will be a 3-dimensional sub-manifold $\mathcal{S}$ such that the restriction $\gamma_{\mu\nu}$ of the spacetime metric $\gl ab$ to the   $\mathcal{S}$ is degenerate. It can be easily shown  (see \eq{nullgeo1} of Appendix) that the normal $\ell^a$ to $\mathcal{S}$ satisfies the null geodesic equation 
$\ell^a \nabla_a \ell_m  \equiv \kappa \ell_m$
with a non-affine parametrization, indicated by a non-zero $\kappa$ which we will call the surface gravity.
Since the normal $\el$ to a null surface is also a tangent, this result also shows that a null surface can be thought of as ``filled by'' (a congruence of) null geodesics. Using this, we can introduce a natural coordinate system adapted to a family of null surfaces in the spacetime.
We  choose one of the coordinates such that $x^3=$ constant correspond to a set of null surfaces with, say, $x^3=0$ on $\mathcal{S}$. Let the intersection of $\mathcal{S}$ with a $x^0=$ constant surface ($\Sigma_t$) of the spacetime
be a 2-dimensional surface $\mathcal{S}_t$
with transverse coordinates $x^A\equiv (x^1,x^2)$ and coordinate basis vectors $\we_A \equiv \partial_A$. We will choose the coordinate system such that
\begin{equation}
\el = \partial_0 +v^A \partial_A = \partial_0 +v^A \we_A; \quad   \ell^a = (1,v^A, 0) 
\label{lcomps}                                                    
\end{equation}  
with  with $\el \w{\cdot} \el =0 = \el \w\cdot \we_A$. 
(That is, we  embed $\mathcal{S}$ in a one-parameter congruence of null hypersurfaces corresponding to $x^3=$ constant and choose the other coordinates $x^0, x^1$ and $x^2$  in such a way that \eq{lcomps} holds.) Clearly, $\el$ has the structure of a convective derivative if we think of $v^A$ as a transverse velocity field.
The line interval, which has nine independent functions, has the form in \eq{metric} of Appendix. In particular the line element on $x^3=$ constant surface is given by:
\begin{equation}
 ds^2 =  q_{AB} (dx^A- v^ A dt ) ( dx^B- v^ B dt )
\label{metric1}
\end{equation} 
The metric on $\mathcal{S}_t$  corresponding to $t=$ constant, $x^3 =$ constant is $q_{AB}$ with a well defined inverse 
$q^{AB}$. The raising and lowering of the uppercase indices $A, B$ etc. in this transverse two-dimensional surface are done using these metrics. 

A more formal way of introducing  this metric is as follows: Given a null surface with the normal $\ell^a$, we first introduce another null vector $k^a$ with $\el \w\cdot \w k=-1$. (For example, in flat spacetime  if $\el= \w e_t+\w e_z$ is the outgoing null vector, then $\w k$ could be $ \w k= (1/2)(\w e_t-\w e_z)$ which is proportional to the ingoing null vector.) 
The metric on the two dimensional surface $\mathcal{S}_t$ orthogonal to this pair  is given by 
the standard relations:
\begin{equation}
 q_{ab} = \gl ab+ \ell_a k_b + \ell_b k_a; \quad q_{ab} \ell^b =0 = q_{ab} k^b
\label{qab}
\end{equation} 
The mixed tensor $q^a_b$ allows us to project quantities onto $\mathcal{S}_t$. 

We are interested in the projection of \eq{ent-func-6} onto $\mathcal{S}$ with $n^a=\ell^a$.   In contrast to timelike or spacelike surfaces, this will now contribute, (i) a term  along $\el $ itself  when we contract \eq{ent-func-6} with $l^a$ (because $\el$ is also tangent to $\mathcal{S}$!) as well as (ii) a projection to the 2-surface $\mathcal{S}_t$ obtained by contracting \eq{ent-func-6} with $q^i_A$. In both, the term involving $\lambda$ will \textit{ not} contribute because $\el^2=0$ and $\ell^a q_{ab}=0$. More formally the projection will give (see \eq{twoterms} of the Appendix) the equations:
\begin{equation}
R_{mn}\ell^m q^n_{a}=8\pi T_{mn}\ell^m q^n_{a}; \quad R_{mn}\ell^m \ell^n=8\pi T_{mn}\ell^m \ell^n
\label{albertpair}
\end{equation} 
In other words, the maximization of the entropy associated with the null vectors lead to these two equations in a natural, direct manner.

Of these, the second equation in \eq{albertpair} involving$R_{mn} \ell^m\ell^n$ will give the familiar Raychaudhuri equation (which we will \textit{not} discuss  in detail) while the first one with  $R_{mn} \ell^m q^n_a$ will lead to an equation that looks like the Navier-Stokes equation --- which will be of our interest in this paper.

To obtain the latter in a physically transparent way, we only have to consider the $A$ th component (in the transverse direction) of the vector $R_{mj}\ell^m$. This can be expressed in terms of the covariant derivative of $\el$ by the standard identity
\begin{equation}
R_{mA}\ell^m = R_{\mu A }\ell^\mu = \nabla_\mu (\nabla_A \ell^\mu) - \partial_A (\nabla_\mu \ell^\mu) 
\label{rma}
\end{equation}
where the first equality arises from \eq{lcomps}. (Recall that in our notation, Greek letters cover $0,1,2$ while uppercase Latin letters cover $1,2$ in the transverse direction.)
To rewrite this as a Navier-Stokes equation,
we need to re-express the derivatives $\nabla_A \ell^\mu$ and $\nabla_\mu \ell^\mu$ in terms of  quantities related to the extrinsic geometry of $\mathcal{S}$. We will now briefly overview the concepts involved in the extrinsic geometry of a null surface delegating details to the Appendix.

\subsection{Extrinsic geometry of a null surface}
\label{sec:nulltech}

The extrinsic geometry of the null surface is determined by the derivative of the normal $\nabla_\mu\el$ along the tangential directions of $\mathcal{S}$. 
Because $\el\w\cdot \nabla_\mu \el = (1/2)\partial_\mu \el^2 =0$ the covariant derivative of $\el$ along vectors tangent to $\mathcal{S}$ is orthogonal to $\el$ and hence is tangent to $\mathcal{S}$. Therefore $\nabla_\alpha \el$ is a vector which can be expanded using the coordinate basis $\we_\mu = \partial_\mu $ on $\mathcal{S}$. (This basis is made of $ \partial_\mu = (\partial_0, \partial_A)=(\el - v^A\we_A, \we_B) = \we_\mu$.) Writing this expansion with a set of coefficients (called Weingarten coefficients) $\ch \alpha\beta$ we have
\begin{equation}
\nabla_\alpha  \el \equiv \ch \beta\alpha \partial_\beta =\ch \beta\alpha \we_\beta ; \qquad \nabla_\alpha \ell^\beta =   \ch \beta\alpha
\label{defchi}             
\end{equation} 
From $\ell^a \nabla_a \ell^\beta = \ell^\alpha \nabla_\alpha \ell^\beta = \ch \beta\alpha \ell^\alpha = \kappa \ell^\beta$, 
it follows that $\ell^\alpha$ is an eigenvector of the matrix $\ch\beta\alpha $ with the eigenvalue $\kappa$.
The coefficients $\ch \beta\alpha$ determine the extrinsic geometry of the null surface we are interested in.
We will now study the different components of the $3\times 3$ matrix $\ch\beta\alpha $. 

To do this, we define the quantity
$\Theta_{\alpha\beta} \equiv \gamma_{\mu\beta} \ch \mu\alpha $ the components of which are given by
\begin{equation}
 \Theta_{\alpha\beta} =\gamma_{\mu\beta} \ch \mu\alpha =  \we_\alpha \w\cdot \nabla_\beta \el =  \nabla_\beta ( \el\w\cdot \we_\alpha) - \el\w\cdot \nabla_\beta \we_\alpha =- \ell_j \Gamma^j_{\alpha\beta}
\label{thetachris}
\end{equation}  
where we have used $\el\w\cdot \we_\alpha =0$. This quantity is clearly symmetric ($\Theta_{\alpha\beta} =\Theta_{\beta\alpha} $) because Christoffel symbols are symmetric. Further, from the result  $\ell_b \nabla_a \ell^b=0$, we have the identity
$ \Theta_{\alpha\beta} \ell^\beta =0$ which, on expansion, using $\ell^\beta = (1, v^A)$, gives $ \Theta_{\alpha 0} = -\Theta_{\alpha B}v^B$.
This implies $ \Theta_{0A}= \Theta_{A0} = -\Theta_{AB}v^B$
which, in turn, leads to $\Theta_{00} = -\Theta_{0B} v^B=v^A v^B \Theta_{AB}$. Thus all components of $\Theta_{\alpha\beta}$ can be determined in terms of the three components of $\Theta_{AB}$ and the metric coefficients $v^A$.

The components $\Theta_{\alpha\beta}$ have  a direct geometrical significance as the Lie derivative along $\el$ of the metric $q_{ij}$ on $\mathcal{S}$. 
Using the   restriction of the 4-dimensional Lie derivative of $\gl ij$ to $\mathcal{S}_t$, we get 
 \begin{equation}
 \Theta_{AB} = \frac{1}{2} {\mathrm{\pounds}}_{\el} q_{AB} 
\label{lie}
\end{equation} 
 in the adapted coordinate system. This structure is similar to the standard formula $K_{\mu\nu}=-(1/2){\mathrm{\pounds}}_{\w n} h_{\mu\nu}$ for the extrinsic curvature of a $t=$ constant spacelike surface
in the $(1+3)$ decomposition. In fact, evaluating \eq{lie} in the coordinate system adapted to $\mathcal{S}$ we get a result very similar to the familiar one for $K_{\mu\nu}$, (see e.g., Eq. (12.21) of ref. \cite{gravitation}): 
\begin{equation}
\Theta_{AB}=\frac{1}{2}\left( D_Av_B+D_Bv_A+ \frac{\partial q_{AB}}{\partial t}\right)
\label{imp1} 
\end{equation} 
When $q_{AB}$ is independent of $t$, we see that $ \Theta_{AB}$ is essentially the shear tensor of a velocity field $v_A$. \textit{This is the key reason why the projected equations can be interpreted as the Navier-Stokes equation.}
 
We will also need to define three more quantities $\omega_\alpha=(\omega_0,\omega_A)$ in terms of the Weingarten coefficients  by 
\begin{equation}
 \omega_\alpha \equiv \ch 0\alpha   = \nabla_\alpha \ell^0 =  \Gamma^0_{j\alpha}\ell^j
\label{chiomega}
\end{equation} 
The utility of $\Theta_{AB}$ and $(\omega_0,\omega_A)$ arises from the fact that
  we can determine \textit{all} the components of $\ch \alpha\beta$ from $\Theta_{AB}, \omega_\alpha$ and the metric coefficients. 
 Among the components of $\ch \alpha\beta$ we already have  the direct relations $\ch 00 = \omega_0, \ch 0A = \omega_A$.  As for the remaining components $\ch A0, \ch AB$, note that:
$\Theta_{0B} = \ch \mu 0 \gamma_{\mu B} = \ch 00 \gamma_{0B} + \ch A0 q_{AB}$ giving
\begin{equation}
 \ch A0 = (\Theta_{0B} - \ch 00 \gamma_{0B}) q^{BA} = -( \Theta_{BC} v^C - \omega_0 v_B) q^{BA}
\label{comp1}
\end{equation} 
Similarly we have
\begin{equation}
 \ch CA   = ( \Theta_{AB} - \ch 0A \gamma_{0B})  q^{BC} = ( \Theta_{AB} + \omega_A v_B) q^{BC}
\label{comp2}
\end{equation} 
So the covariant derivative $\nabla_\mu \el$ of the normal vector $\el$ --- and thus the extrinsic geometry
of $\mathcal{S}$ --- is completely characterized by the set $(\Theta_{AB} = -\ell_m\Gamma^m_{AB}, \omega_0 = \ell^m \Gamma^0_{m0}$, $\omega_A = 
\ell^m \Gamma^0_{mA})$.  Confined to the transverse components, \eq{comp2} also gives the relation: 
\begin{equation}
 \nabla_A \ell_B =  \chi_{AB} = [\Theta_{AB} + \omega_B v_A] 
\label{trans}
\end{equation} 
which will turn out to be useful.

Being a symmetric tensor, $\Theta_{AB}$ can be expressed in terms  of its irreducible parts, viz. the trace $\theta \equiv \Theta^A_A$ and the trace-free part $\sigma^A_B$ by 
\begin{equation}
\Theta^A_B=\sigma^A_B+(1/2)\delta^A_B\theta 
\label{tracesep}
\end{equation} 
In the adapted coordinates, $\theta $ is given by the expression
\begin{equation}
 \theta = \frac{\partial}{\partial t} \ln \sqrt{q} + D_A v^A
\label{imp2}
\end{equation} 
and is related to the divergence of $\el$ through the relation
\begin{equation}
\nabla_a \ell^a = \nabla_\mu \ell^\mu = \nabla_A \ell^A + \nabla_0\ell^0 = (\theta + v_A \omega^A) + \omega_0 = \theta + \ell^\mu \omega_\mu \equiv \theta + \kappa
\label{divell}
\end{equation} 
In arriving at the third equality we have used \eq{trans} and in arriving at the first and the  fourth equality we have used \eq{lcomps}. The last relation arises because $\ch 0\mu \ell^\mu = \kappa \ell^0 = \kappa = \omega_\mu \ell^\mu$. 

\subsection{The Navier-Stokes equation in the boosted inertial frame}
\label{sec:NSinboost}

Using these results, we are in a position to  rewrite \eq{rma} in terms of the variables related to the extrinsic geometry of the null surface. Using \eq{defchi} and \eq{divell} we can rewrite \eq{rma} as:
\begin{equation}
R_{mA}\ell^m = R_{\mu A }\ell^\mu = \nabla_\mu \ch \mu A - \partial_A(\kappa+\theta)
\label{rma1}
\end{equation}
Let us consider this expression  around any given event in a frame in which  the Christoffel symbols vanish but their derivatives do not. We will assume that the metric coefficients become constant and have  their usual diagonal form  except for keeping $v^A\neq 0$ but  a constant. (That is,   we are working with a Lorentz-boosted local inertial frame; we will comment on this choice later on in Sec. \ref{sec:rescale}.) In that case, we can write $\nabla_\mu \ch \mu A$ in \eq{rma1} as:
\begin{eqnarray}
\nabla_\mu \ch \mu A &=& \partial_\mu \ch \mu A = \partial_0 \ch 0A + \partial_B \ch BA = \partial_0 \omega_A + \partial_B (q^{BC} \chi_{CA})\nonumber\\
&=& \partial_0 \omega_A +\partial_B\left[ q^{BC} \left( \Theta_{CA} + \omega_A v_C\right)\right]=\partial_0\omega_A + \partial_B\Theta^B_A +v^B \partial_B \omega_A
\end{eqnarray}
Each equality is obtained by discarding terms involving Christoffel symbols but retaining terms containing their derivatives;  we have also used \eq{chiomega} and \eq{trans}. (Note, for example, that $\omega_\mu$ is proportional to  Christoffel symbols and hence vanishes but its derivative need not.) 
Thus  we get
\begin{eqnarray}
R_{mA} \ell^m &=& ( \partial_0 +v^B \partial_B ) \omega_A + \partial_B \left( \sigma^B_A + \frac{1}{2} \theta \delta^B_A\right) - \partial_A (\kappa+\theta)\nonumber\\
&=& (\partial_0 +v^B\partial_B ) \omega_A + \partial_B\sigma^B_A - \partial_A\left(\kappa +\frac{1}{2}\theta\right)
\label{firstresult}
\end{eqnarray}
where we have used \eq{tracesep}. The first equation in \eq{albertpair} now becomes:
\begin{equation}
(\partial_0 +v^B\partial_B ) \omega_A + \partial_B\sigma^B_A - \partial_A\left(\kappa +\frac{1}{2}\theta\right)=8\pi T_{mA} \ell^m 
\label{NS1} 
\end{equation} 
Again note that $\sigma^B_A$ vanishes in the local inertial frame but its derivative $\partial_B\sigma^B_A$ will not.
Rewriting \eq{NS1} in the form
\begin{equation}
(\partial_0 +v^B\partial_B )\left( -\frac{\omega_A}{8\pi}\right) = \frac{1}{8\pi}\partial_B\sigma^B_A -\frac{1}{16\pi}\partial_A\theta 
- \partial_A\left(\frac{\kappa}{8\pi}\right) -  T_{mA} \ell^m 
\label{NS2} 
\end{equation}
we see that \eq{NS2} has the \textit{exact form}  of a Navier-Stokes equation for a fluid with (i) momentum density $-\omega_A/8\pi$, (ii) pressure $(\kappa/8\pi)$, (iii) shear viscosity coefficient $\eta=(1/16\pi)$. (Note that in the conventional NS equation the viscous tensor $2\eta\sigma^A_B+\xi\delta^A_B\theta$ is defined with an extra factor 2 for shear viscosity.) (iv) bulk viscosity coefficient  $\xi=-1/16\pi$ and (v) an external force $F_A= T_{mA} \ell^m$.  The NS will also have term $\theta \omega_A$ which vanishes in the local frame because $\theta=0$ but, of course, we can formally add this term (which is numerically zero) to \eq{NS2} to complete the structure; but  the derivative $\partial_A\theta$ is non-zero, which allows us to uniquely determine the bulk viscosity term. 

For this interpretation to be strictly valid, it is also necessary that $\sigma^B_A$ has the form of a trace-free shear tensor built from the velocity field $v_A$. From \eq{imp1} we see that this is true provided $(\partial q_{AB}/\partial t)=0$ when we have:
\begin{equation}
 \partial_B\sigma^B_A= \frac{1}{2}\partial_B[(\partial^B v_A+\partial^A v_B)-\delta^B_A (\partial_Cv^C)]
\end{equation} 
In the literature,  \eq{NS2} or its analogues are often called as  Navier-Stokes equation even when $(\partial q_{AB}/\partial t)$ is non-zero. This is --- strictly speaking --- incorrect especially because the metric $q_{AB}$ has no fluid dynamical interpretation in general. We shall come back this point in the next section.

It should also be noted that, in the conventional NS equation, the momentum density that is transported by the fluid is usually collinear with velocity. In our case, the $\omega_A$ and $v_A$ are in general unrelated and --- in fact --- the physical meaning of $\omega_A$ is unclear for a general null surface. (It does have a physical meaning in the context of the event horizon of Kerr black hole, for which the formalism was originally delevoped; in that context, $\omega_A$ can be related to the angular momentum of the black hole.)

It is clear that $\eta/\xi=-1$ also makes any interpretation in terms of standard viscous dissipation unnatural for a \textit{general} null surface \cite{comment}. In the literature, this interpretation  has been provided in situations when teleological evolution is acceptable especially  in the context of black hole event horizon. More importantly, the ratio between shear viscosity and entropy density $s=1/4$ of the null surface is given by $\eta/s=1/4\pi$, when we use the fact that any null surface can be attributed an area density of entropy which is (1/4) in Einstein's theory. Similar results are known in much more complicated situations arising from string theory but they presumably have the basis in \eq{NS2}.

For completeness we will also state the Raychaudhuri equation arising from the second equation in \eq{albertpair} which, in the boosted local inertial frame reads (see \eq{rai} of Appendix and note that $\Theta_{mn} =0=\theta$ in this frame):
\begin{equation}
  R_{mn} \, \ell^m \ell^n  =  - (\partial_0 +v^B\partial_B )\theta =8\pi T_{mn} \, \ell^m \ell^n.
\label{rai1}
\end{equation}
which relates the evolution of $\theta$ to the energy flux across the null surface. In the context of local Rindler frame, the right hand side of \eq{rai1} can be interpreted as proportional to the heat flux making this an entropy balance equation. We shall briefly comment on this interpretation in Sec. \ref{sec:entdenfn}.  

\subsection{The Damour-Navier-Stokes equation in arbitrary frame}
\label{sec:DNS}

The entire analysis analysis can be repeated in an arbitrary coordinate system to obtain a generalized Navier-Stokes equation, first obtained by Damour \cite{damourthesis}.  This is derived in the Appendix (see \eq{rmneqn}) and has the form:
\begin{equation}
 R_{mn} \, \ell^m q^n_{a} =q^m_{a} {\mathrm{\pounds}}_{\el} \Omega_m + \theta\, \Omega_a - D_a \left( \kappa + \frac{\theta}{2} \right) +   D_m \sigma^m_{a}=8\pi T_{mn} \, \ell^m q^n_{a}
\label{rmneqn1}
\end{equation} 
where the various terms are defined below. The metric $q^a_b$ is given by \eq{qab}; the $\Omega_m$ is the projection
$\Omega_m=q^n_m\omega_n=\Omega_n-\kappa k_n$ where $\omega_n=\ell^j\nabla_j k_n$; note that, since $k_A=0$, we have $\Omega_A=\omega_A$ as far as the transverse components,  which we are interested in, are concerned. The $\Theta_{mn}$ is the projection of the covariant derivative:
$\Theta_{mn}=q^a_mq^b_n\nabla_a \ell_b$. The $D_i$ is the covariant derivative operator defined using the metric $q_{ab}$ in the two-dimensional subspace, while ${\mathrm{\pounds}}_{\el}$ is the Lie derivative with respect to $\el$. 

A comparison of \eq{rmneqn1} with \eq{firstresult} shows that --- while most terms have an one-to-one correspondence, with suitable projections replacing the operation of taking transverse components --- there are  \textit{two crucial differences} which we will now discuss: 

(i) In the literature one usually sees the interpretation of $\sigma^m_{a}$ as the viscous shear tensor of a fluid with a velocity field $v_a$ without bothering about its internal structure. But, as we have already mentioned, this tensor contains an extra term involving the time derivative of the transverse metric, $(\partial q_{ab} /\partial t)$ which has no direct fluid dynamical interpretation. Similarly, the bulk viscosity term involving $\theta$ contains the time derivative $(\partial/\partial t) \ln \sqrt{q}$. (See \eq{imp1} and \eq{imp2}; of course, $(\partial q_{ab}/\partial t) =0$ in the  frame we are working with, but its derivative $\partial_m (\partial q_{ab}/\partial t) $
does not vanish, which is what leads to the difficulty). To get out of this difficulty, one needs to make certain choices for the metric  and coordinate system so that $q_{ab}$ is independent of $t$. (In some other contexts one can avoid this term by rescaling $\el$ or by working in a perturbative series for the metric; we shall briefly comment on these possibilities in Sec. \ref{sec:rescale}.)
There is very little discussion of this issue in the literature and the term involving $\sigma^m_a$ is usually called shear viscosity term without worrying about its internal structure.

(ii) The second, and probably more crucial, difference is the following: 
The DNS equation in a general coordinate system, given by \eq{rmneqn1},
contains the \textit{Lie derivative} with respect to $\el$ while the standard fluid dynamical, non-relativistic,  NS equation contains the \textit{convective derivative} with respect to $\el$. This difference can be explicitly verified by expanding out \eq{rmneqn1} in the coordinate system adapted to the null surface in which $\el=\partial_0+v^A D_A$ (see \eq{expDNS} in Appendix): 
\begin{equation}
 ( \partial_0 + v^B D_B) \omega_A + \omega_B\,  D_A v^B + \theta \omega_A + D_B \sigma^B_A -D_A \left(\kappa + \frac{1}{2} \theta\right)=8\pi T_{mn} \ell^n q^m_A 
\label{expDNS1}
\end{equation}
The key difference is second term in the left hand side, $\omega_B\,  D_A v^B$, involving the \textit{derivative of the velocity field} which is absent in the standard NS equation, making the two equations structurally different.(This term vanishes in the local inertial frame since $\omega_A$ vanishes in that frame, making the structure of DNS equation identical to NS equation.) More formally, the convective derivative term  $D_t \Omega_a = q^n_a \ell^m \nabla_m \Omega_n$ defined with correct projections, is related to Lie derivative term by:
\begin{equation}
 D_t \Omega_a = q^n_a \ell^m \nabla_m \Omega_n = q^m_a {{\mathrm\pounds}}_{\el} \Omega_n - \Theta^m_a \Omega_m
\end{equation} 
Using this we can rewrite \eq{rmneqn1} as:
\begin{equation}
D_t \Omega_a+\Theta^m_a \Omega_m + \theta \, \Omega_a -  D_a \left( \kappa + \frac{\theta}{2} \right) + D_m \sigma^m_{a}=8\pi T_{mn} \, \ell^m q^n_{a}
\label{rmneqn2}
\end{equation} 
which clearly brings in an extra term, which is the second term in the left hand side.

Our \eq{rmneqn1} and \eq{expDNS1} agree with the original results obtained by Damour [see e.g., Eq. (I.55) of ref. \cite{damourthesis} or Eq. (15b) of ref \cite{damourGR}]. In ref. \cite{damourthesis}, Damour also quotes the standard NS equation (see Eq. (I.47) of ref \cite{damourGR} ) which, of course, differs by the term involving the derivative of the velocity from the Eq. (I.55) of the same work. Similar difference exists in ref.\cite{damourstring} between the DNS equation [Eq.2.67] and the NS equation [Eq. 2.68]. This result also has been obtained in the membrane paradigm approach in ref. \cite{pricethorn} and our \eq{rmneqn2} matches with Eq. (2.14) of ref. \cite{pricethorn}; the second term in the left hand side of their Eq (2.14) is the extra term $\Theta^m_a \Omega_m$ of \eq{rmneqn2} making it different from the NS equation. (Price and Thorne, however, does \textit{not} call this equation Navier-Stokes equation, probably because of the extra term, and refers to it as Hajicek equation.) Since our derivation in the Appendix is modeled exactly as in ref. \cite{ericjose}, it, of course, matches with their equations in Section (6.3).
This review, as well as many other papers in the literature (like, for e.g., \cite{elingoz}), call our \eq{rmneqn1} or \eq{expDNS1} the DNS equation even though there is an extra term. It is true that the Lie transport arises more naturally than convective transport in curved spacetime and one may think of it as a natural generalisation; but the extra term in a PDE makes its structure (and solutions) very different and hence a careful distinction needs to be made.

We have, however, shown that one can indeed think of DNS equation as identical to the NS equation provided we view it in the (boosted) local inertial frame. Since $\omega_a$ is proportional to the Christoffel symbols, the extra term vanishes in the local inertial frame making this interpretation easy. This procedure actually has a broader domain of relevance which I will now describe. 

\section{Fluid dynamics of null surface and the issue  of viscous dissipation}
\label{sec:DNSdisc}

In any generally covariant theory, one tries to express the equations in a form valid in arbitrary coordinate systems in order to exhibit the symmetries explicitly. While this is certainly important to check the diffeomorphism invariance of the theory, it has two shortcomings in certain contexts: 

(a) The metric coefficients will depend on the spacetime coordinates both in curvilinear coordinates as well as in a curved spacetime. Principle of equivalence implies that one cannot meaningfully ask ``how much'' of this dependence arises due to the choice of the coordinates and ``how much'' arises due to the curvature of spacetime.
When we are interested in the \textit{structural aspects} of the theory, the spacetime dependence of the metric due to the choice of coordinates acts as an extra baggage which one would like to avoid. Working in the local inertial frame allows us to do this.

(b) The emergent paradigm of gravity suggests that one needs to accept an intrinsic observer dependence in all thermodynamical variables. In particular, one can attribute an entropy, temperature and related dynamics to local Rindler horizon as perceived by the corresponding Rindler observer in the spacetime even though a freely falling observer will not attribute  any  thermodynamical features  to the same null surface (\cite{tpdialogue,marolf}; also see Sec. 4.4 of ref. \cite{rop}). But, while we use the observer dependent thermodynamics to obtain the field equations of gravity in this approach, the final result should be (and is) independent of the Rindler observers and must be generally covariant. Therefore while projecting the field equations to a null surface and interpreting them as NS equation of a fluid living on that null surface,   it is crucial to ask which part of the field equation survives in the freely falling frame. Our analysis shows that the terms involving (i) the derivatives of the stress tensor, (ii) the derivative of the pressure and the (iii) external momentum flux term remain non-zero in the freely falling frame because they involve \textit{derivatives} of the Christoffel symbols. This is in spite of the fact that the viscous stress tensor and the pressure themselves  vanish in the freely falling frame. What is relevant for the  equation to be interpreted as the NS equation is the existence of a viscous \textit{force} on the  fluid, arising from the \textit{gradient} of the viscous tensor (and pressure), rather than the viscous tensor (or pressure) itself. Of course, this is very counterintuitive compared to normal fluid mechanics. If flow of water exhibits viscosity, it will also have non-zero viscous tensor in the same frame of reference! But here, we are merely calling some combination of Christoffel symbols as viscosity tensor (see \eq{thetachris}) and their derivatives involving certain combinations of Ricci tensor as viscous force. Obviously, the former can vanish without the latter vanishing in a local inertial frame.

This feature has important implications for the characterization of viscous dissipation in the current context. In the conventional fluid dynamics based on NS equation, the viscous dissipation will be proportional to terms involving $\sigma_{AB} \sigma^{AB}$ and $\theta^2$ and --- in usual fluid mechanics  --- $\sigma_{AB} \sigma^{AB}, \theta$ and $\partial_B \sigma^{AB}$ will all be nonzero. In our case $\sigma_{AB}$ and $\theta$ vanish in the freely falling frame and \textit{the inertial observers in spacetime will not see any dissipation}. (The Raychaudhuri equation  in arbitrary coordinates --- \eq{rai} in the Appendix --- has these quadratic term in the viscous tensor which disappear in \eq{rai1} valid in the  locally inertial frame.) This is reassuring since we do not probably expect a continuous, \textit{observer independent}, dissipation to take place in spacetime. But the force on the viscous fluid, which depends on the gradient $\partial_B \sigma^{B}_A$ and $\partial_A\theta$, does not vanish even in the freely falling frame showing that this force has an observer independent existence.  It might appear, at first, paradoxical that a viscous force term exists for the fluid but no viscous dissipation. But, as we said above, algebraically this is no more paradoxical than the fact that effects due to curvature involving derivatives of Christoffel symbols can be present locally even when the Christoffel symbols vanish at a point. The result  obtained above is just a translation of this well-known fact in the language of null surface dynamics. In fact the real paradox would be if there is an observer independent viscous dissipation in spacetime!

There have been attempts in the literature to interpret the equations for the fluid on the null surface, especially the Raychaudhuri equation, as describing  entropy production (\cite{damourthesis,damourGR,pricethorn}; for more recent work based on emergent approach to gravity, see e.g., \cite{eiling}). The early work (e.g. \cite{damourthesis,damourGR,pricethorn} leads to  an entropy production that is acausal (or teleological, as it is often called)  involving the entire future history of the null surface. In this context, we need to distinguish between two different physical situations. This teleological feature is probably acceptable in the context in which DNS equation was originally derived and applied, viz., to black hole event horizons. Event horizons are in any case teleological even in their definition and hence this is not a surprise. 

But the DNS equation is applicable to \textit{any} null surface in the spacetime, including patches of the local Rindler horizons. In the emergent paradigm of gravity one relies heavily on local Rindler horizons which are ordinary null surfaces in spacetime and not event horizons corresponding to some specific \textit{solutions}  of field equations. In fact, \textit{to obtain the field equations} from a suitably defined entropy density of spacetime, one uses the local Rindler horizons as \textit{off-shell} constructs in the theory and hence they cannot be linked to horizons arising in on-shell solutions. But since, in our derivation of DNS we never assigned any special property to the null surface in question,  the resulting equations are applicable as equally to a local Rindler horizon in a given spacetime as to an event horizon of a black hole obtained as specific solution. This is why we need to carefully distinguish generic features of interpretation which are valid for \textit{any} null surface from some of the features which might have limited validity in the specific context of black hole horizon. The dissipation and generation of entropy with accompanying irreversible thermodynamics may be acceptable in the case of physical processes involving black hole horizons, say, but such dissipation  is difficult to interpret  in terms of local Rindler horizons. So, again, it is probably welcome that $\sigma_{AB} \sigma^{AB}$ and $\theta^2$ vanish in the freely-falling frame but the NS equation for the fluid remains valid. From this point of view, it  seems better to interpret the equations in the freely falling frame.

Incidentally, it does not seem to have been emphasized in the literature that Einstein's equations have a very simple form in the freely falling frame. Using the exact expression for Ricci scalar
\begin{equation}
 R_{ab} = \frac{1}{\g} \partial_m(\g\, \Gamma^m_{\phantom{m}ab}) - \partial_a\partial_b(\ln \g)- \Gamma^m_{\phantom{m}na}\Gamma^n_{\phantom{m}bm}
\end{equation}
we find that, in the locally inertial frame Einstein's equations take the form:
\begin{equation}
 R_{ab} = \partial_m\Gamma^m_{\phantom{m}ab} - \partial_a\partial_b \g=8\pi (T_{ab}-\frac{1}{2}\eta_{ab}T)
\label{albertFF}
\end{equation}
In fact projecting \eq{albertFF} to a null surface (when $\eta_{ab}T$ term will not contribute) we can re-derive our NS equation.
 Since all the terms in the left hand side of \eq{albertFF} correspond to specific combinations of  curvature tensor components we see that the DNS equation in the boosted inertial frame essentially provides mapping between curvature components and the relevant fluid variables.  This aspect will be explored in a separate publication \cite{tpprogress}.

\subsection{Spacetime dependent boosts and rescaling of the null normal}
\label{sec:rescale}

In obtaining \eq{NS1} we have purposely used a boosted local inertial frame in which all the metric coefficients take the standard Cartesian form except $g_{0A} = - v_A$. One could have, of course, repeated the derivation with $v_A$ [and its derivative which, being  proportional to Christoffel symbols, are anyway zero] set to zero but with the second derivatives of $v_A$ (which contribute to the viscous force term in $\partial_B\sigma^B_A$) remaining non-zero. We kept the $v_A$ non-zero in the original derivation just to exhibit the nature of convective derivative and to contrast it with the Lie derivative. Physically, one can think of non-zero $v_A$ as arising because of a transverse velocity for the observers (e.g., local Rindler observers with a constant velocity drift parallel to the Rindler horizon). 

The derivation of DNS equation does not pre-suppose any special form of the metric.
In certain contexts, one can choose the metric as well as the normalization of $\el$ in such a way that the extra term in the Lie derivative is cancelled out. This arises because the different terms in the DNS equation scale differently when we rescale the null vector $\el$ by a spacetime dependent factor $\gamma (x)$. (Being a null vector, $\el$ has no natural normalization, without us making some additional assumptions.) This transformation changes the components of $\ell^i$ to the form ($\gamma, 0, \gamma v^A$) suggesting the interpretation of a spacetime dependent Lorentz boost with a $\gamma$ factor. One can easily show that when $\el \to \el'= \gamma \el $ the $\Theta_{ab}, \sigma_{ab} $ and $\theta$ scales by a factor $\gamma$ and the vector $\w k $ goes to $\gamma^{-1} \w k$. But $\kappa$ and $\Omega_A$ change inhomogeneously as: 
\begin{equation}
 \kappa' = \gamma ( \kappa + \nabla_{\el} \ln \gamma); \quad \Omega'_A = \Omega_A + D_A \ln\gamma
\label{rescale}
\end{equation} 
 Because of \eq{rescale}, the DNS equation does not retain its form  under such a rescaling. One can, by hand, choose the value of $\gamma$ to cancel out the extra term in the Lie derivative thereby reducing the equations to the standard N-S form.  (This has been done, for example, in ref. \cite{elingoz} with a different motivation.) 

A similar mathematical situation arises in  a more general context of a  set of transformations which are usually used in the literature. Consider a metric written in a coordinate system with $\gl 0\alpha =0$ so that the line element is 
\begin{equation}
 ds^2 = -F^2 dt^2 + f_{\alpha\beta}dx^\alpha dx^\beta
\label{lel}
\end{equation}  
We will now substitute in this metric $dt=u_a dx^a$ where $u_a$ has the components $u^a = \gamma(1,v^\alpha)$. If we take
$v^\alpha$ to be \textit{constant} with $\gamma = (1-v^2)^{-1/2}$, then such a transformation is equivalent to a Lorentz boost and, of course, leaves the physics invariant. But when $v^\alpha$ and $\gamma $ are functions of spacetime coordinates, the transformation changes the character of geometry. (Without additional constraints, $dt= u_a(x) dx^a $ will not even be integrable and $dt$ will not be an exact differential; however, one can certainly use this ``rule" to modify the metric.) Straightforward algebra now shows that the resulting metric can be expressed in the standard (1+3) form as:
\begin{equation}
 ds^2 = - N^2 dt^2 + h_{\alpha\beta} (dx^\alpha + N^\alpha dt)(dx^\beta + N^\beta dt)
\label{newmetric}
\end{equation} 
where 
\begin{eqnarray}
 N^2 &=& F^2 \gamma^2 ( 1-F^2 \gamma^2 v^2)^{-1}; \quad h_{\alpha\beta} = f_{\alpha\beta} - F^2 \gamma^2 v_\alpha v_\beta\nonumber\\
N^\mu &=& F^2 \gamma^2 v^\mu ( 1-F^2 \gamma^2 v^2)^{-1}; \quad N_\mu = F^2 \gamma^2 v_\mu
\label{drifttrans}
\end{eqnarray}
Obviously, such a modification of the metric is not a symmetry of the theory when $u^a(x)$ are spacetime dependent functions; for example, if the original metric in \eq{lel} satisfied Einstein's equations with some source, the modified metric in \eq{newmetric} will not be a solution unless very special conditions are satisfied. We will provide a general discussion of these transformations in a future publication \cite{tpprogress} but will point out one simple feature here which is relevant to our discussion. 

Notice that if we consider the lowest order change, linear in the boost velocity, it is essentially the generation of the cross term with $N_\mu \approx F^2 v_\mu$. This has the effect of generating  an extra term to the extrinsic curvature of the $t=$ constant surface given by
\begin{equation}
 \delta K_{\alpha\beta} = \frac{1}{2F} [D_\alpha (F^2 v_\beta) + D_\beta(F^2 v_\alpha)]
\end{equation} 
where $D$ is the covariant derivative based on the spatial 3-metric.
When the original coordinate system was synchronous with $F=1$,  this extra term has the form
\begin{equation}
 \delta K_{\alpha\beta} = \frac{1}{2} [D_\alpha  v_\beta + D_\beta v_\alpha]
\end{equation} 
Further, if we treat one of the spatial coordinates, say, $x^1$ differently from the two transverse coordinates $x^A = (x^2, x^3)$, then the $\Theta_{AB}$ of the null surfaces orthogonal to null vectors in $x^0-x^1$ plane will pick up a term of the form $\delta \Theta_{AB} = (1/2) [D_A  v_B + D_B v_A]$ (see \eq{formula1} in the Appendix). This, in turn, corresponds to the addition of an extra term
\begin{equation}
\delta \sigma_{AB} = \frac{1}{2} [(D_A  v_B + D_B v_A) - q_{AB}(D_C v^C)] 
\label{viscoust}
\end{equation} 
to the trace-free part of $\Theta_{AB}$ 
which has the form of the shear viscosity tensor. This analysis shows how a drift velocity in the transverse direction to a null surface can lead to the viscous stress tensor of the correct form. Several discussions of this topic in the literature are essentially special cases of the transformation in \eq{drifttrans} with very specific choices for $F, f_{\alpha\beta}$ and $u_a$. One is led to an NS equation in all these contexts essentially because of the result in \eq{viscoust}.

As a simple illustration of the same effect, consider a metric of the form 
\begin{equation}
ds^2 = - N^2 dt^2 + \frac{dx^2}{N^2} + \delta_{AB}( dx^A - v^A dt)( dx^B - v^B dt)
\end{equation} 
which could represent 
the local Rindler frame when $v^A=0$ and $N(x) \propto x$ with the surface $x=0$ being null and acting as the local Rindler horizon.
Since such a metric represents flat spacetime, it obviously satisfies source-free Einstein's equation. If $v^A\neq 0$ \textit{but constant}, the metric represents  a boosted frame in which the Rindler observer has a constant drift velocity $v^A$ in the transverse direction. When $v^A  = v^A (x, x^A)$ the metric is still static but will not, in general, satisfy source-free Einstein's equations. This introduces a spacetime dependent boost velocity with non-zero shear. For a general $N(x)$ the null surfaces with translational symmetry in the $y,z$ directions are given by $f(t,x) = $ constant where 
\begin{equation}
f(t,x) = - t + \int\frac{dx}{N^2} 
\end{equation} 
The normal to this 
null surface  
 can be taken to be $\ell_a = N^2 (-1, N^{-2}, 0,0)$. (The overall normalisation, as we have discussed, is not unique but this form is based on \eq{stdlk} of the Appendix, which has some level of naturalness.)  A direct calculation shows that, for these null surfaces, we have 
\begin{equation}
\Theta_{AB} = \frac{1}{2} [\partial_{A} v_{B}-\partial_{B} v_{A}]
\end{equation} 
with $\Omega_A = (1/2) \partial_x v_A$
and $\kappa = \partial_x N^2$. It is  obvious that $\Theta_{AB}$ has the form of a shear tensor for a fluid with velocity $v^A$. So the null surface fluid equations will pick up a viscous tensor arising from this shear. This is essentially what happens even in more general contexts.

Finally, it may be noted that the validity of DNS equation on a null surface forms the basis for the structural similarity noticed between gravity and fluid mechanics in the context of string theory motivated approaches in AdS etc (see e.g., \cite{adspap}). Once the field equation is reduced to the DNS form (or its generalization in a higher dimensional AdS context), one can attempt to solve the equation --- which is an issue we have not addressed --- and obtain the metric systematically using a long wavelength approximation of the fluid mechanics. This would involve expanding the metric in a series  $\gl ab= \gl ab^0+\epsilon \gl ab^1+\epsilon^2\gl ab^2 ....$ indexed by a book-keeping parameter $\epsilon$ and, simultaneously introducing a derivative expansion by rescaling the coordinates by $x^i \to \epsilon x^i$ in the argument.  This will allow one to  solve for the metric order by order in the long wavelength approximation exactly as in the case of fluid mechanics. 

\section{Conclusions}
The similarity between gravitational field equations and the Navier-Stokes equation in the context of null surfaces was known (at least to the relativity community) for several decades now. However, it had not found any specific utility except possibly as a conceptual tool in membrane paradigm until recently when similar results, inspired by string theory, attracted attention to this subject. In this backdrop, the results of this paper have the following significance and implications:

\subsection{Structure of spacetime entropy density functional}\label{sec:entdenfn}

In all the previous approaches to the DNS equation, one starts from the gravitational field equations (which would have been obtained by some standard field theoretic procedure), projects them onto a null surface and reinterprets the resulting equation as the DNS equation. This is somewhat unsatisfactory conceptually since one would have expected a \textit{fluid dynamical} equation to arise from an entropy maximization principle or something similar. Given the fact that Einstein's field equation itself has clear thermodynamical interpretation, it is obvious that DNS equation itself must be derivable in a direct manner from entropic considerations. We have achieved this goal in this paper by reinterpreting the previous result \cite{rop,aseemtp} of obtaining the field equations from extremising entropy density of spacetime. 

It is conceptually satisfying that the extremum condition for the entropy density of spacetime  (directly related to null surfaces because they can act as local Rindler horizons) given by \eq{ent-func-6} has a natural interpretation when projected to a null surface. As \eq{albertpair} shows, such a projection --- involving contractions with either $\ell^a$ or $q^a_b$ --- makes the Lagrange multiplier term in \eq{ent-func-6} disappear and gives a direct relation between Ricci tensor and energy momentum tensor. 

It may also be noted that the gravitational entropy density  --- which is the integrand $s_{grav}\propto ( -P_{ab}^{cd} \D_c\ell^a\D_d\ell^b)$ in \eq{Sgrav} --- obeys the relation:
\begin{equation}
 \frac{\partial s_{\rm grav}}{\partial ( \nabla_c \ell^a)}\propto (- P^{cd}_{ab} \nabla_d \ell^b) \propto (\nabla_a \ell^c  - \delta^c_a \nabla_i \ell^i)
\label{sgravder}
\end{equation} 
where the second relation is for Einstein's theory. This term is analogous to the more familiar object $t^c_a = K^c_a - \delta^c_a K$ (where $K_{ab}$ is the extrinsic curvature) that arises in the (1+3) separation of Einstein's equations. (More precisely, the appropriate projection to 3-space leads to $t^c_a$.) This combination can be interpreted as a surface energy momentum tensor in the context of membrane paradigm because $t_{ab}$ couples to $\delta h^{ab}$ on the boundary surface when we vary the gravitational action ( see, e.g., eq.(12.109) of \cite{gravitation}). In fact, one obtains the results for null surfaces as a limiting process from the time-like surfaces (usually called stretched horizon) in the  of membrane paradigm \cite{pricethorn}. Equation~(\ref{sgravder}) shows that  the entropy density of spacetime is directly related to  $t^c_a$ and its counterpart in the case of null surface.   

This term also has the interpretation as the canonical momentum conjugate to the spatial metric in (1+3) context and \eq{sgravder} shows that the entropy density leads to a similar structure. This will form the basis for generalising the DNS equation to \LL\ models in a future publication \cite{tpprogress}.

Further, the \textit{functional} derivative  of the gravitational entropy in \eq{Sgrav} has the form, in any \LL\ model:
\begin{equation}
 \frac{\delta S_{\rm grav}}{\delta \ell^a} \propto \mathcal{R}_{ab}\ell^b \propto J_a
\end{equation} 
Previous work \cite{rop,surfaceprd,tpdialogue} has shown that the current $J_a= 2\mathcal{R}_{ab} \ell^b$ plays a crucial role in interpreting gravitational field equations as entropy balanced equations. In the context of local Rindler frames, when $\ell^a$ arises as  a limit of the time-like Killing vector in the local Rindler frame, $J_a$ can be interpreted as the Noether (entropy) current associated with the null surface. In that case, the generalization of the two projected equations in \eq{albertpair} to \LL\ model will read as
\begin{equation}
 J_a \ell^a = \frac{1}{2} T_{ab} \ell^a \ell^b; \quad J_aq^a_m = \frac{1}{2}T_{ab}\ell^a q^b_m
\end{equation} 
which relate the gravitational entropy density and flux to matter energy density and momentum flux. The second equation in the above set becomes the DNS equation in the context of Einstein's theory. All these results, including the DNS equation,  will have direct generalization to \LL\ models which can be structured using the above concepts \cite{tpprogress}.
We again see that all these ideas find a natural home in the emergent paradigm. 

\subsection{Comparison of DNS and NS equations}\label{sec:compare}

We have used this occasion to clarify some issues related to the interpretation of DNS equation as describing a viscous fluid. As noted in Section \ref{sec:DNS} there are, in general,  two crucial differences between DNS equation and standard NS equation. The first is the appearance of time derivatives of the transverse metric $q_{ab}$ in the definition of shear viscosity tensor because of which it cannot be interpreted properly in general. In fact this term has no hydrodynamical interpretation except in very specific contexts or in a perturbative series. 
One can, however, avoid this difficulty by choosing specific class of metrics and coordinate systems. 

The second difference has to do with the appearance of Lie derivative rather than the convective derivative in the momentum transport equation. As I argued, this requires one to work in the boosted local inertial frame for proper interpretation. Somewhat surprisingly, these two differences have not been explicitly discussed in published literature, to the extent I know.

\subsection{Dissipation: a new level of observer dependence}

The existence of  viscosity in the fluid equation raises questions related to possible dissipational effects. I am not comfortable with the notion of  continuous, \textit{observer independent}, dissipation in the spacetime and have stressed the fact  that the dissipational terms involving $\sigma_{ab}\sigma^{ab}$ and $\theta^2$ vanish in the boosted inertial frame which I consider to be appropriate for the interpretation of DNS equation, in view of the  comments in Sec. \ref{sec:DNS} and Sec. \ref{sec:compare}. As pointed out earlier, in such a frame,
the derivatives of the viscous tensor (related to spacetime curvature) are non-zero while the viscous tensor (proportional to Christoffel symbols) vanish. Of course, an observer who is \textit{not} freely falling will perceive non-zero dissipational terms while a freely falling observer will not. This may sound paradoxical but observer dependence in the thermodynamic description of horizons is very well known and need not cause any (new) surprise. 

A general framework incorporating the observer dependence of thermodynamic variables and providing a translation table for physical phenomena perceived by observers in different states of motion is currently lacking.  This aspect  deserves further investigation.

\section*{Acknowledgements}

I thank  C.Eling, D. Kothawala and S.Liberati for useful discussions.

\section*{Appendix: Null surfaces}
Consider a 4-dimensional spacetime manifold $\mathcal{M}$ with a metric $g_{ab}$. 
A null surface  in this spacetime is a 3-dimensional sub-manifold $\mathcal{S}$ such that the restriction $\gamma_{\mu\nu}$ of the spacetime metric $\gl ab$ to   $\mathcal{S}$ is degenerate, i.e., there exist vectors $v^\mu$ in $\mathcal{S}$ such that $\gamma_{\mu\nu} v^\mu =0$.  (Recall that we use the mostly positive signature; the 
Latin indices, $a,b...$ go over $0-3$ in $\mathcal{M}$ while the Greek indices $\alpha,\beta, ...$ go over the  coordinates in a 3-dimensional null surface $\mathcal{S}$ with signature $(-, +, +)$. When we restrict ourselves to the two-dimensional, spatial sub-manifold of this null surface, we will use uppercase Latin indices $A,B,...$. Our results are easily generalizable to a $D-$dimensional manifold but we will stick with 4 dimensions for notational simplicity.) 

The normal to $\mathcal{S}$ is  a null vector field $\ell^a$ in the spacetime which can be written in the form $\ell_a = f\partial_a \phi$ where $f$ and $\phi$ are scalars in the spacetime and $\phi$ is constant on $\mathcal{S}$. Straightforward algebra using the fact $\nabla_i \nabla_j \phi =\nabla_j \nabla_i \phi$ shows that 
\begin{equation}
 \ell^a \nabla_a \ell_m = \left( \ell^i\partial_i \ln f \right) \ell_m \equiv \kappa \ell_m
\label{nullgeo1}
\end{equation} 
Therefore the rescaled vector $L^m \equiv f^{-1} \ell^m$ is  null and satisfies the geodesic equation with affine parameterization; i.e., $L^i \nabla_i L^j =0$. Hence a null surface can be thought of as ``filled by'' (a congruence of) null geodesics.
Equation (\ref{nullgeo1}) shows that $\ell^a$ also satisfies the geodesic equation but with a non-affine parameterization if $\kappa$ is non-zero. When the null surface corresponds to a black hole horizon in an asymptotically flat spacetime, there is a natural choice for $\ell^a$ such that $\kappa$ can be identified with the surface gravity of the black hole horizon. We shall continue to use this terminology and call $\kappa$ in \eq{nullgeo1} as the surface gravity of $\mathcal{S}$ even when it  depends on the choice of normalization for the null vector field. In the case of $\mathcal{S}$ being a local Rindler horizon, one can relate $\kappa$ to the acceleration of the congruence of Rindler observers which one is considering; the arbitrariness in the normalization translates into the arbitrariness in the choice of the acceleration for the Rindler observers.

The fact that the null surface is spanned by the null geodesics allows us to introduce a natural coordinate system adapted to a family of null surfaces in the spacetime as follows: 
We  choose one of the coordinates such that $x^3=$ constant correspond to a set of null surfaces with, say, $x^3=0$ on $\mathcal{S}$. 
Let the intersection of $\mathcal{S}$ with a $x^0=$ constant surface ($\Sigma_t$) of the spacetime
be a 2-dimensional surface $\mathcal{S}_t$
with coordinates $x^A\equiv (x^1,x^2)$ and coordinate basis vectors $\we_A \equiv \partial_A$.
 At any point $\mathcal{P}$ in $\Sigma_t$ there will be one future pointing null direction orthogonal to the 2-surface $\mathcal{S}_t$. We choose $\el$ at $\mathcal{P}$  to be in this direction with $\el\w \cdot \we_A =0$ on $\mathcal{S}_t$. We can now erect a coordinate system in the neighborhood of $\Sigma_t$ by the following choice: (a) The coordinates $x^1, x^2, x^3 $ are taken to be constant along the geodesics starting from each point $\mathcal{P}(x^1,x^2,x^3)$ in $\Sigma_t$ in the direction of $\ell^a (x)$; (b) $x^0$ is chosen to be  the affine parameter distance along these geodesics with $\lambda =t$ on $\Sigma_t$. 
 
 In such a coordinate system, $\el = \partial/\partial x^0$ so that $\ell^a = \delta^a_0$ and the condition $\el^2=0$ translates to $\gl 00 =0$. Further, the geodesic condition with affine parametrization gives 
\begin{equation}
0=\ell^b \nabla_b \ell^a= \nabla_0 \ell^a = \Gamma^{a}_{00} = \frac{1}{2} \gu ab (2 \partial_0 \gl 0b - \partial_b \gl 00)  = g^{ab} \partial_0 g_{0b}                                                         
\end{equation} 
requiring $\partial_0 \gl 0b =0$ along each geodesic. But initially, on $\Sigma_t$ we have $\gamma_{0A} = \el \w\cdot \we_A=0$
and $\gamma_{00} =0$ because $\el $ is orthogonal to the basis vectors $\we_A$ as well as to itself. This requires $(\gamma_{00},\gamma_{0A}) $ to vanish all along the geodesic. 
The line element will now take the form:
\begin{equation}
 ds^2 = - N^2 dt^2 + \left(\frac{M}{N} dx^3+\epsilon Ndt \right)^2 + q_{AB} (dx^A + m^A dx^3) ( dx^B + m^B dx^3)
\label{premetric}
\end{equation}
 with $\epsilon = \pm 1$. The $x^3=$ constant surfaces are null with the line element $ds^2=q_{AB}dx^Adx^B$ because in the (degenerate) metric $\gamma_{\mu\nu}$ (with $\mu,\nu=0,1,2$) the coefficients $\gamma_{0\mu}$ vanish: i.e.,  $\gamma_{00} = \gamma_{0A} =0$. For example, when  $N=M=1;q_{AB}=\delta_{AB};m^A=0$ we recover the usual null coordinates $x^3=z+t$ or $x^3=z-t$ for the two choices of $\epsilon$.
 
In this construction of coordinates, we have embedded $\mathcal{S}$ in a one-parameter congruence of null hypersurfaces corresponding to $x^3=$ constant and the other coordinates $x^0, x^1$ and $x^2$ are constructed in such a way that $\el = \partial/\partial {x^0}$ is a null geodesic field in the neighborhood of $\mathcal{S}$. 
While this is always possible it is often advantageous to use a slightly less restrictive coordinate system in which the geodesic condition is relaxed but the coordinate $x^3$ is constant on the null hypersurfaces of interest.
This can be achieved by allowing for non-zero $\gamma_{00}$ and $\gamma_{0A}$. 
 In such a case, the   coordinates can be chosen such that the line interval becomes:
\begin{equation}
 ds^2 = - N^2 dt^2 + \left(\frac{M}{N} dx^3+\epsilon Ndt \right)^2 + q_{AB} (dx^A- v^ A dt + m^A dx^3) ( dx^B- v^ B dt + m^B dx^3)
\label{metric}
\end{equation} 
The metric on $\mathcal{S}_t$  corresponding to $t=$ constant, $x^3 =$ constant is $q_{AB}$ with a well defined inverse 
$q^{AB}$. The raising and lowering of the uppercase indices $A, B$ etc. in this transverse two-dimensional surface are done using these metrics. We also have $\g = M ({\rm det}\  q_{AB})^{1/2}$.
The null vector field will now be  $\el = \partial_0 +v^A \partial_A = \partial_0 +v^A \we_A$ with components $\ell^a = (1,v^A, 0)$ with $\el \w{\cdot} \el =0 = \el \w\cdot \we_A$. Clearly, $\el$ has the structure of a convective derivative if we think of $v^A$ as a transverse velocity field.

From the form of the metric in \eq{metric} it can be explicitly verified that $\gamma_{\mu\nu} \ell^\nu =0$ where the Greek letters run over $0,1,2$ on the null surface $\mathcal{S}$ with $x^3=$ constant. Therefore the 3-metric on $\mathcal{S}$ is indeed degenerate. This fact requires us to be careful in characterizing the extrinsic geometry of $\mathcal{S}$ since raising and lowering of indices will not be possible with a degenerate metric. As outlined in the main text, this can be done by expanding  $\nabla_\alpha \el$  using the coordinate basis $\we_\mu = \partial_\mu $ on $\mathcal{S}$. 
For many purposes it is convenient to have a more formal approach in terms of tensorial objects defined in the 4-dimensional spacetime and their projections onto $\mathcal{S}_t$. We shall now indicate how this can be done along the lines described in the excellent review, ref. \cite{ericjose}. 

We will begin
by introducing the standard $(1+3)$ foliation of the spacetime with the normals $\w n = - N \w d t$  to $\Sigma_t$ where $N$ is the lapse function.  Let $\w s$ be a unit normal to a set of time-like surfaces such that $\w n \w\cdot \w s =0$. 
We can now define 
 two null vector fields by
\begin{equation}
 \el = N ( \w n + \w s); \qquad \w k = (1/2N) (\w n - \w s)
 \label{stdlk}
\end{equation} 
The primary vector field we are interested in is $\el$ while $\w k$ is an auxiliary null vector field with $\el \w \cdot \w k = -1$ which is useful for the study of extrinsic geometry. (Of course, $\el$ and $\w k$ can be introduced without $\w n$ and $\w s$ but this allows a natural normalization.) We can now define a metric $q_{ab}$ on the 2-dimensional surface $\mathcal{S}_t$ orthogonal to the $\w n$ and $\w s$ through standard relations:
\begin{equation}
 q_{ab} =\gl ab + n_a n_b - s_a s_b = \gl ab+ \ell_a k_b + \ell_b k_a; \quad q_{ab} \ell^b =0 = q_{ab} k^b
\end{equation} 
The mixed tensor $q^a_b$ allows us to project quantities onto $\mathcal{S}_t$. We can also define another projector orthogonal to $k^b$ by the definition 
$\pip db = \delta^d_b  + k^d \ell_b$
which has the properties 
\begin{equation}
\pip ab \ell^b = \ell^a ; \quad \pip ab k^b =0; \quad \pip ab \ell_a =0; \quad \pip ab k_a =k_b.                                                                                    \end{equation} 
We can now introduce the Weingarten coefficients as the projection of the covariant derivative $\nabla_d \ell^a$ by the definition \begin{equation}
\ch ab \equiv \pip db \nabla_d \ell^a = \nabla_b \ell^a + \ell_b ( k^d \nabla_d \ell^a)
\label{defchi1}                                                                                                                                                                                                                        \end{equation} 
which has the properties
\begin{equation}
 \ch ab \ell^b \equiv \kappa \ell^a; \ \chi_{ab} k^b = 0; \ \chi_{ab} \ell^a = 0; \  \chi_{ab} k^a  \equiv - \omega_b= - \ell^j \nabla_j k_b
\end{equation} 
where the surface gravity $\kappa$ is defined through the relation $\ell^j \nabla_j \ell_i = \kappa \ell_i$ and $\omega_a $
through the last equality. The only non-trivial result is $ \chi_{ab} k^a=- \ell^j \nabla_j k_b$ which can be proved (see e.g., Eq. (5.40) of ref. \cite{ericjose})  by working out the components in the adapted coordinate system.
Note that $\omega_a$ satisfies the relations $\omega_a \ell^a = \kappa$ and $\omega_a k^a =0$.
 We next define $\Theta_{ab}$ by projecting $\chi_{mb}$ to $\mathcal{S}_t$. We get, on using $\ell^m \chi_{mb} =0 $ and $k^m \chi_{mb} = - \omega_b$, the result:
\begin{equation}
 \Theta_{ab} = q^m_a \chi_{mb} = \chi_{ab} + k_a \ell^m \chi_{mb} + \ell_a k^m \chi_{mb} = \chi_{ab} - \ell_a \omega_b
\end{equation} 
which is essentially \eq{trans} expressed in the four-dimensional notation with suitable projection tensors.
Using \eq{defchi} we see that 
\begin{equation}
 \Theta_{ab} = \Theta_{ba} = \nabla_b \ell_a + \ell_a k^i \nabla_i \ell_b - \ell_b \omega_a = q^m_a  q^n_b \nabla_m \ell_n
\label{deftheta}
\end{equation} 
This result shows that $\Theta_{ab}$ is a natural projection of the covariant derivative $\nabla_m \ell_n$ onto the surface $\mathcal{S}_t$ and, obviously, $\Theta_{ab}\ell^b=0=\Theta_{ab}k^b$.
The trace of $\Theta_{ab}$,  denoted by $\theta$, is given by
\begin{equation}
 \Theta^a_a = \theta = \nabla_a l^a - \kappa
\label{deftheta1}
\end{equation} 
which corresponds to \eq{divell}.
It is also convenient  to define a similar projection of $\omega_a$ by $\Omega_b \equiv q^a_b \omega_a$. We have
 \begin{equation}
 \Omega_b \equiv q^a_b \omega_a    = - q^a_b (k_m \ch ma ) = \omega_b  - \kappa k_b ( k_m \ell^m) = \omega_b + \kappa k_b                                                                                                                                                                                                                                                                                                                                                                                                                                                                                                                     \end{equation} 
For computational purposes it is often convenient to relate $\Theta_{ab}, \Omega_{i}$ etc. to more familiar quantities defined using the standard (1+3) decomposition of the metric. It can be shown that (see Section 10.2 of ref. \cite{ericjose}) the following results hold which are often useful for explicit computation:
\begin{equation}
 \Theta_{ab} = N(D_m s_n - K_{mn}) q^m_a q^n_b; \quad \theta = N (D_\alpha s^\alpha + K_{\alpha \beta} s^\alpha s^\beta - K)
 \label{formula1}
\end{equation} 
\begin{equation}
 \kappa = \ell^m \nabla_m \ln N + s^\alpha D_\alpha N - N K_{\alpha\beta} s^\alpha s^\beta; \quad \Omega_a = D_a \ln N - K_{mn}s^m q^n_a
\end{equation} 

These results again allow  us to express the projection of Einstein's equations onto $\mathcal{S}_t$. 
To do this one begins with the standard relation
$\nabla_m \nabla_a \ell^m - \nabla_a \nabla_m \ell^m = R_{ma} \, \ell^m $ 
and substitute for $\nabla_a \ell^m$ using \eq{deftheta} and for $\nabla_m \ell^m$ using \eq{deftheta1} repeatedly. This leads, after some straightforward algebra, to the relation 
\begin{eqnarray}
  R_{ma} \, \ell^m & = & \nabla_m \Theta^m_{a} + \ell^m \nabla_m \omega_a + (\kappa + \theta) \omega_a  - \nabla_a (\kappa + \theta) - \Theta_{am} k^n \nabla_n \ell^m \nonumber \\
  & & - \left(\omega_m k^n \nabla_n \ell^m + \nabla_m k^n \, \nabla_n \ell^m  + k^n \nabla_m \nabla_n \ell^m \right) \ell_a
 \label{expansion1}
\end{eqnarray}
The projection of $R_{ma} \ell^m$ using $\pip na$ separates into two terms given by
\begin{equation}
R_{mn}\, \ell^m\Pi^n_{a} =  - R_{mn}\ell^m \ell^n \, k_{a}  +   R_{mn}\ell^m q^n_{a} . 
\label{twoterms}                               
\end{equation} 
As mentioned in the main text,  we now see that there is a projection along $\el $ itself  and a projection to the 2-surface $\mathcal{S}_t$. Of these, the $R_{mn} \ell^m\ell^n$ will give the familiar Raychaudhuri equation while $R_{mn} \ell^m q^n_a$ will lead to an equation which looks similar to Navier-Stokes equation. The Raychaudhuri equation can be obtained very easily by contracting \eq{expansion1} with $\el$  and simplifying the result. We get 
\begin{equation}
  R_{mn} \, \ell^m \ell^n  = - \Theta_{mn} \Theta^{mn}  + \kappa\theta - \ell^m \nabla_m \theta .
\label{rai}
\end{equation} 
It is again conventional to separate out the trace of $\Theta_{mn}$ and define $\sigma_{mn} = \Theta_{mn} - (1/2)q_{mn} \theta$ so that we can write 
$\Theta_{mn} \Theta^{mn} = \sigma_{mn} \sigma^{mn}+(1/2) \theta^2 $.

The derivation of the Navier-Stokes like equation is more complicated.  Contracting \eq{expansion1} with $q^a_b$ leads to the expression 
\begin{eqnarray}
R_{mn} \, \ell^m q^n_{a} & = &  q^n_{a} \nabla_m \Theta^m_{n}  + q^n_{a} \ell^m \nabla_m \omega_n + (\kappa + \theta) \Omega_a  -  D_a (\kappa + \theta) \nonumber \\
& &  - \Theta_{am} k^n \nabla_n \ell^m  .  
\label{nveqn}
\end{eqnarray}
with $q^n_a \nabla_n (\kappa+\theta) \equiv  D_a(\kappa+\theta)$ where $ D_a$ is the
covariant derivative  defined using the metric on $\mathcal{S}_t$. The first and the last terms on the right hand side can also be combined in terms of $ D_a$ using the relation:
\begin{eqnarray}
  D_a \Theta^a_b &\equiv& q^i_j q^k_b \nabla_i \Theta^j_k = \left( \delta^i_j + \ell^i k_j + \ell_j k^i \right) q^k_b  \nabla_i \Theta^j_k\nonumber\\
&=& q^k_b  \nabla_i \Theta^i_k -  \Theta^j_b \nabla_i ( \ell^i k_j + \ell_j k^i ) 
=  q^k_b  \nabla_i \Theta^i_k - \Theta^j_b ( \ell^i\nabla_i k_j +k^i\nabla_i  \ell_j )\nonumber\\
&=&  q^k_b  \nabla_i \Theta^i_k - \Theta^j_b ( k^i \nabla_i\ell_j +\Omega_j)
\end{eqnarray}
In obtaining this result we have repeatedly used the orthogonality condition $\Theta^i_a \ell^a =0=\Theta^i_a k^a$ and the definitions $\ell^n\nabla_n k_m = \omega_m, \ q^m_j \omega_m = \Omega_j$. Therefore we can combine the first and last terms in \eq{nveqn} as
\begin{equation}             
q^n_{a} \nabla_m \Theta^m_{n} - \Theta^m_{a}  k^n \nabla_n \ell_m   = D_m \Theta^m_{a} + \Theta^m_{a}  \Omega_m  . 
\end{equation}
In the second term in \eq{nveqn} we introduce the Lie derivative of $\Omega_n$ through the expansion 
\begin{eqnarray}
q^n_{a} \ell^m \nabla_m \omega_n & = & q^n_{a} \ell^m \nabla_m (\Omega_n- \kappa k_n) = q^n_{a} ( \ell^m  \nabla_m \Omega_n
- \kappa \ell^m \nabla_m k_n) \nonumber \\
&= &  q^n_{a} \left( {\mathrm{\pounds}}_{\el} \Omega_n - \Omega_m \nabla_n \ell^m  - \kappa \omega_n\right) \nonumber \\
&= &  q^n_{a}  {\mathrm{\pounds}}_{\el} \Omega_n - \Theta_a^{m} \Omega_m - \kappa \Omega_a , 
\end{eqnarray}
Combining all these together we get 
\begin{eqnarray}
R_{mn} \, \ell^m q^n_{a}  &=& q^m_{a}  {\mathrm{\pounds}}_{\el} \Omega_m + \theta \, \Omega_a -  D_a (\kappa + \theta)  +   D_m \Theta^m_{a} \nonumber\\
&=&q^m_{a}  {\mathrm{\pounds}}_{\el} \Omega_m+ \theta \, \Omega_a -  D_a \left( \kappa + \frac{\theta}{2} \right) +   D_m \sigma^m_{a}
\label{rmneqn}
\end{eqnarray} 
where we have again set $\Theta^m_n = \sigma^m_n + (1/2) \delta^m_n \theta$.

This result can be re-expressed in several different ways. For example, the convective derivative of $\Omega_n$  has the form 
\begin{equation}
 D_t \Omega_a = q^n_a \ell^m \nabla_m \Omega_n = q^m_a {{\mathrm\pounds}}_{\el} \Omega_n - \Theta^m_a \Omega_m
\end{equation} 
using which we can express \eq{rmneqn} in terms of the convective derivative. Alternatively, we can express it  in the coordinate system adapted to $\mathcal{S}$, in which the projection of the Lie derivative has the form 
\begin{equation} 
q^m_{B}  {\mathrm{\pounds}}_{\el} \Omega_m = \pdov{\Omega_A}{t}+ v^B \,  D_B \Omega_A + \Omega_B \,  D_A v^B .
\end{equation}
where we have used $\el = \partial_0 +v^A\we_A$. Then \eq{rmneqn} becomes, on using $\Omega_A=\omega_A$ for transverse components, because $k_A=0$, 
\begin{equation}
R_{mn} \ell^n q^m_A = ( \partial_0 + v^B\  D_B) \omega_A + \Omega_B\,   D_A v^B + \theta \omega_A +  D_B \sigma^B_A - D_A \left(\kappa + \frac{1}{2} \theta\right) 
\label{expDNS}
\end{equation}
 This is the form  which was used in the main text.


\begin{thebibliography}{99}
\bibitem{rop} 
Padmanabhan T.,   \textit{Rep. Prog. Phys.},  \textbf{73} (2010) 046901, [arXiv:0911.5004].

\bibitem{daviesunruh}
Davies P C W (1975)  \textit{J. Phys.} A \textbf{8}  609--616; 
Unruh W G  (1976) \textit{Phys. Rev.} D \textbf{14}  870.


\bibitem{surfaceprd}
T. Padmanabhan, \textit{Mod. Phys. Lett.} \textbf{A 25}, 1129 (2010), [arXiv:0912.3165];
T.Padmanabhan, \textit{Phys. Rev.}, \textbf{D 81}, 124040 (2010), [1003.5665].

\bibitem{wald} 
Wald R. M.,  \textit{Phys. Rev. D}  (1993), {\bf  48}  3427, [gr-qc/9307038];
Iyer V.  and R. M. Wald, (1995), \textit{Phys. Rev. D} {\bf 52}  4430, [gr-qc/9503052].


\bibitem{aseemtp} 
Padmanabhan T.,   (2008), \textit{Gen.Rel.Grav.},   \textbf{40}, 529-564 [arXiv:0705.2533];
Padmanabhan T.  and Paranjape A., (2007), \textit{Phys.Rev.} D, \textbf{75}, 064004 [gr-qc/0701003].






 \bibitem{damourthesis} 
T. Damour, (1979), 
\textit{Quelques propri\'et\'es m\'ecaniques, \'electromagn\'etiques,
thermo\-dy\-na\-mi\-ques et quantiques des trous noirs},
Th\`ese de doctorat d'\'Etat, Universit\'e Paris 6. (available at http://www.ihes.fr/~damour/Articles/)

\bibitem{damourGR}
T. Damour, (1982), \textit{Surface effects in black hole physics},
in \textit{Proceedings of the Second Marcel Grossmann Meeting on General
Relativity}, Ed. R. Ruffini, North Holland , p. 587.

\bibitem{gravitation}
T.Padmanabhan (2010) \textit{Gravitation: Foundations and Frontiers}, Cambridge University Press, UK.


\bibitem{comment} As an aside, we may mention that the condition of for a $2\times 2$ symmetric tensor $S^{AB}$ (with 3 independent components) to be expressible as a viscous stress-tensor of a 2-dimensional velocity field (with 2 independent components) with viscosity coefficients $\eta,\xi$ is $\partial_A\partial_B S^{AB}=(1/2)[1+(\eta/\xi)]S$. Using this result and the (1+3) decomposition of of Einstein's equations, one can show that $(\eta/\xi)=-1$ when the fluid interpretation is possible for Einstein's equations.

\bibitem{damourstring}
T. Damour and M. Lilley, (2008), \textit{String theory, gravity and experiment}, 
arxiv:0802.4169.


\bibitem{pricethorn}
R.H. Price and K.S. Thorne, (1986), 
\textit{Phys. Rev.}  {\bf D 33}, 915. 

\bibitem{ericjose}
E. Gourgoulhon and Jose Luis Jaramillo, (2006),\textit{ Phys.Rept.},\textbf{ 423}, 159. [gr-qc/0503113].


\bibitem{elingoz}
C. Eling and Y. Oz, (2010), \textit{JHEP}, 1002:069. [arXiv:0906.4999].


\bibitem{tpdialogue} 
Padmanabhan T 2009 \textit{A Dialogue on the Nature of Gravity} [arXiv:0910.0839]

\bibitem{marolf} 
Marolf D,  Minic D and  Ross S 2004 \textit{ Phys.Rev.}  \textbf{D69}  064006.


\bibitem{eiling}
C.  Eiling, R. Guedens, T. Jacobson, \textit{Phys. Rev. Letts.} \textbf{96}, 121301 (2006); 
C. Eiling, \textit{JHEP} \textbf{11}, 048 (2008);
G. Chirco and S. Liberati, \textit{Phys.Rev.},\textbf{ D81}, 024016, (2010); 
G. Chirco, C. Eling, S. Liberati, [arXiv:1011.1405].


\bibitem{tpprogress}
T.Padmanabhan (2010), work in progress.


\bibitem{adspap}
See e.g.,
S. Bhattacharyya, V. E. Hubeny, S. Minwalla and M. Rangamani, \textit{JHEP} \textbf{0802}, 045 (2008);  
S. Bhattacharyya et al., \textit{JHEP} \textbf{0806}, 055 (2008); for a review see
M. Rangamani, \textit{Class.Quant.Grav.}, \textbf{26}, 224003, (2009) 	[arXiv:0905.4352].



\end{thebibliography}
\end{document}